\documentclass[sigconf]{acmart}
\AtBeginDocument{%
  }

\setcopyright{acmlicensed}
\copyrightyear{2025}
\acmYear{2025}
\acmDOI{XXXXXXX.XXXXXXX}
\acmConference[MM '25]{Proceedings of the 33th ACM International Conference on Multimedia}{October 27--31, 2025}{Lisbon, Portugal}
\acmISBN{XXX-X-XXXX-XXXX-X/25/10}

\usepackage{pifont}
\usepackage{bm}
\usepackage[ruled,vlined]{algorithm2e}
\usepackage{booktabs}

\begin{document}

\title{MMC: Iterative Refinement of VLM Reasoning via MCTS-based Multimodal Critique}

\author{Shuhang Liu}
\authornote{Equal contributions.}
\author{Zhenrong Zhang}
\authornotemark[1]
\affiliation{%
  \institution{NERC-SLIP, University of Science and Technology of China}
  \city{}
  \country{}
}

\author{Pengfei Hu}
\author{Jiefeng Ma}
\affiliation{%
  \institution{NERC-SLIP, University of Science and Technology of China}
    \city{}
  \country{}
}


\author{Jun Du}
\author{Qing Wang}
\authornote{Corresponding author.}
\affiliation{%
	\institution{NERC-SLIP, University of Science and Technology of China}
  \city{}
\country{}
}


\author{Jianshu Zhang}
\author{Quan Liu}
\affiliation{%
	\institution{IFLYTEK Research}
	  \city{}
	\country{}
}


\author{Jianqing Gao}
\author{Feng Ma}
\affiliation{%
	\institution{IFLYTEK Research}
	  \city{}
	\country{}
}


\renewcommand{\shortauthors}{Trovato et al.}
\renewcommand\footnotetextcopyrightpermission[1]{} 
\settopmatter{printacmref=false} 
\newcommand{\supgain}[1]{\textsuperscript{\textcolor{green!60!black}{$\uparrow$#1}}}
\newcommand{\suploss}[1]{\textsuperscript{\textcolor{red!70!black}{$\downarrow$#1}}}
\begin{abstract}
  
  Visual language models (VLMs) have demonstrated strong performance across diverse multimodal reasoning tasks but still face challenges such as hallucinations, resulting in incorrect reasoning outcomes. Inspired by recent research on external feedback mechanisms in large language models (LLMs), we propose a multimodal actor-critic framework to enhance VLM reasoning capabilities. Specifically, the actor model generates step-by-step reasoning paths based on image and text inputs, while the critic model evaluates these reasoning paths and provides corrective feedback. The actor model iteratively refines its reasoning based on the feedback until the reasoning outcome is deemed satisfactory by the critic model. To reduce reliance on costly manual annotations, we introduce an automated method for constructing multimodal critique datasets. By leveraging Monte Carlo Tree Search (MCTS), we systematically guide the actor model to explore diverse reasoning paths. To obtain critique data for correcting erroneous reasoning steps, we prompt an annotator model to compare pairs of reasoning paths diverging from a shared ancestor node—one leading to a correct conclusion and the other to an incorrect one. This approach enables us to construct the MMC ($\textbf{M}$CTS-based $\textbf{M}$ultimodal $\textbf{C}$ritique) dataset, upon which we further develop a comprehensive training and inference pipeline. Extensive experiments conducted on several public benchmark datasets and mainstream VLMs demonstrate that our approach significantly improves the performance of VLM on complex multimodal reasoning tasks, underscoring its effectiveness and wide applicability.

\end{abstract}

\begin{teaserfigure}
	\includegraphics[width=\textwidth]{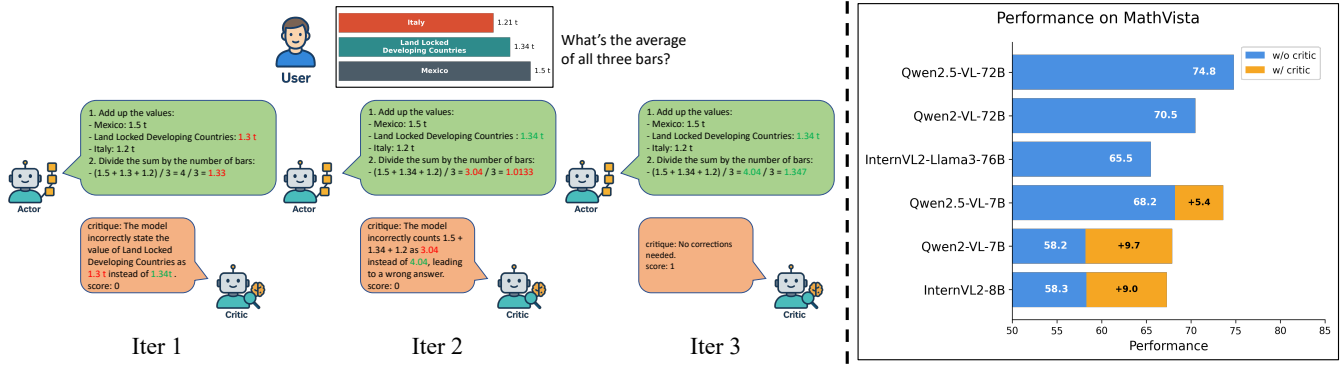}
	\caption{Overview of the multimodal actor-critic framework (left), and its performance across multiple VLMs on the MathVista benchmark (right).}
	\label{fig:overview}
\end{teaserfigure}



\settopmatter{printfolios=true}
\maketitle

\section{Introduction}
Visual Language Models (VLMs) have overcome the limitations of unimodal methods, enabling a more comprehensive and contextual understanding of the world~\cite{2024-CoRR-Multimodal_In-Context_Learning,2024-Artif-A_survey_on_ml}. They have shown excellent performance in various fields, such as multimodal dialogue~\cite{2023-CoRR-TouchStone} and visual question answering~\cite{2024-CoRR-VLM-VQA}. However, VLMs also face unique challenges that differ from those of unimodal models. A prominent issue is visual hallucination, where the model generates responses primarily driven by the embedded parametric knowledge with its Large Language Model (LLM) components rather than based on actual visual content~\cite{2024-CVPR-Hallusionbench,2024-CoRR-Hallucination-VLM}. Additionally, hallucinations can also arise within the LLM module itself, leading to errors even when the model is processing purely text information~\cite{2023-ACM-Hallucination-LLM}.  Early methods primarily relied on the "direct prediction" strategy~\cite{2024-CVPR-Chat-UniVi,2024-CVPR-VILA}, while simple, has proven inefficient for tasks requiring logical reasoning~\cite{2024-CoRR-CoT-VLM}. The Chain-of-Thought (CoT) method~\cite{2022-NeurIPS-CoT} mitigates this limitation by prompting the model to explicitly generate intermediate reasoning steps. However, VLMs are still prone to errors and hallucinations during reasoning, leading to incorrect conclusions~\cite{2023-CoRR-CoT-Faithfulness,2023-NeurIPS-CoT-Verification,2023-NeurIPS-CoT-Unfaithful}. Therefore, enhancing the reasoning capabilities of VLMs remains an important research direction.


To address these challenges, OpenAI o1~\cite{2024-OpenAI-o1} has worked on improving model performance on reasoning-intensive benchmarks by increasing inference-time computation, allowing models to match or even surpass human experts' performance.  DeepSeek-R1~\cite{2025-DeepSeek-R1} successfully utilizes reinforcement learning (RL) to promote the self-emergence of advanced cognitive reasoning abilities in LLMs. At the same time, 
many studies employ mechanisms like self-reflection, self-correction, and self-critique to produce longer and more refined reasoning chains~\cite{2023-NeurIPS-Reflexion,2024-ACL-Self-Refinement,2023-ICLR-Self-Correct}.
Building on these advancements, recent efforts in the multimodal domain have explored similar directions, including dynamically refining reasoning trajectories through adaptive prompting~\cite{2024-AtomThink}, stage-wise refining of reasoning annotations~\cite{2024-LLaVA-CoT}, and using Monte Carlo Tree Search (MCTS) to link incorrect and correct paths for learning self-correction~\cite{2025-Agent-R}.

In addition to these efforts in internal reasoning strategies, recent research has explored another paradigm by introducing an external feedback mechanism that decouples the critique generation from reasoning process~\cite{2023-ACL-RL4F,2024-AutoMathCritique,2024-critic-v,2025-CFT}. In this framework, the actor model generates clear, structured reasoning steps, while an independent critic model provides feedback, identifies errors, and guides corrections through iterative interactions. The effectiveness of the critic model depends on the availability of high-quality critique datasets. However, fine-grained manual annotation is costly and time-consuming. To address this issue, existing studies have proposed automated methods for constructing critique datasets, which can be broadly classified into two categories.

The first type of method~\cite{2024-critic-v,2024-AutoMathCritique} involves using large-scale LLMs (e.g., GPT-4o~\cite{2024-gpt4o}) to deliberately modify correct reasoning paths into incorrect ones. By explicitly annotating the error locations and types, the annotator model can generate critiques that directly target these known errors. However, the error paths generated are artificially inserted rather than originating from the base actor model itself, which may limit the critic model's ability to adapt to real reasoning errors. The second type of method~\cite{2024-AutoMathCritique,2025-CFT} involves generating both correct and incorrect reasoning paths through techniques such as repeated sampling by the actor model, and then passing the complete reasoning paths to the annotator model for comprehensive evaluation. The annotator model compares the two full reasoning paths, systematically identifies errors, and generates critiques based on the differences between them. However, this approach faces two key challenges, stemming from the need to compare complete correct and incorrect reasoning paths. First, it requires the annotator model to have strong reasoning and evaluation abilities—often relying on large-scale LLMs like GPT-4o—to ensure accurate error detection and critique generation. Second, by focusing on the full-path comparison, this method struggles to capture fine-grained reasoning errors that occur at specific steps.

To address the concern, we propose an automated process for constructing a multimodal critique dataset without the need for manual annotations. We integrate multiple visual question answering (VQA) datasets, including MathV360k~\cite{2024-EMNLP-Math-LlaVA}, ChartQA~\cite{2022-ACL-ChartQA}, and M\({}^{\mbox{3}}\)CoT~\cite{2024-m3cot}, and guide the actor model through Monte Carlo Tree Search (MCTS)~\cite{2006-MCTS} to systematically explore step-by-step reasoning paths. Unlike previous methods that rely on full-path comparisons, our method takes advantage of the tree structure produced by MCTS. The annotator only needs to compare local differences between diverging branches, making it easier to pinpoint errors and provide step-level corrective feedback. This eliminates reliance on more powerful external models than the base actor model, substantially reducing annotation costs. Moreover, the tree-based organization enables comprehensive coverage of reasoning errors and their corresponding correction strategies at each step.
 Based on this, we construct the MMC ($\textbf{M}$CTS-based $\textbf{M}$ultimodal $\textbf{C}$ritique) dataset. The critic model trained on MMC can effectively identify reasoning errors and generate targeted corrective feedbacks.

Furthermore, we build a multimodal actor-critic framework to enhance the reasoning of VLMs. This framework consists of a actor model and a critic model working in tandem: based on image and text inputs, the actor model generates step-by-step reasoning paths, while the critic model evaluates them and provides targeted feedback. The actor model iteratively refines the reasoning paths based on the feedback until the critic model provides satisfactory feedback. We extensively evaluate this framework on multiple public benchmark datasets and mainstream VLMs. Experimental results show that this method significantly improves VLM performance on complex reasoning tasks and demonstrates good generalization ability and effectiveness. 

In summary, our main contributions are:

	\begin{itemize} 
		
	\item We propose an MCTS-based automated critique generation method, resulting in the MMC dataset, which enables fine-grained supervision for step-level reasoning errors without manual annotations.

	\item We introduce a multimodal actor–critic framework, where the actor iteratively refines its reasoning path based on the corrective feedback from the critic model, enhancing VLM reasoning performance.

	\item We conduct extensive experiments on multiple public multimodal benchmarks across various mainstream VLMs, demonstrating significant performance gains and strong generalization ability.
\end{itemize}

\section{Related Work}
\subsection{Self-Improving Multi-Step Reasoning}
Allocating additional computational resources during the inference process has been shown to significantly enhance the performance of LLMs~\cite{2024-OpenAI-o1}. As an effective approach to reasoning optimization, the CoT method~\cite{2022-NeurIPS-CoT} guides models to reason step by step, mimicking human-like problem-solving by breaking down complex problems into structured intermediate steps. This decomposition improves both the coherence and accuracy of the reasoning process. However, in multi-step reasoning, even small errors in intermediate steps can propagate and lead to significant deviations in the final answer. Traditional approaches~\cite{2024-ICLR-PRM800k} primarily rely on expert-annotated reasoning datasets to enable models to learn step-by-step reasoning patterns. Such annotations are costly and difficult to scale.

To address these challenges, recent research has increasingly explored the use of LLMs to automatically generate high-quality reasoning data and enhance multi-step reasoning capabilities. AtomThink~\cite{2024-AtomThink} leverages GPT-4 to construct reasoning datasets, where short thought-chain amplification extends brief intermediate steps to create more complete reasoning processes, while dynamic prompting transforms datasets containing only final answers into formats with detailed step-by-step reasoning. LLaVA-CoT~\cite{2024-LLaVA-CoT} introduces a four-stage reasoning framework that employs GPT-4 to refine annotations at each stage of the VQA task. During inference, a staged beam search selects the optimal candidate path at each reasoning step, improving the stability and reliability of the reasoning process. Agent-R~\cite{2025-Agent-R} utilizes Monte Carlo Tree Search (MCTS) to generate both correct and incorrect reasoning trajectories, prompting the model to identify and correct the first erroneous step in its reasoning by logically splicing it to the correct path. This approach constructs long-chain multi-step reasoning data with self-reflection, allowing supervised fine-tuning to teach models how to recover from errors and transition toward correct reasoning paths. As a result, this method enhances the robustness and self-correction abilities of multi-step reasoning models.

\subsection{External Feedback for Error Correction}
In addition to the efforts in internal reasoning strategies, recent research has explored another paradigm by introducing an external feedback mechanism.
RL4F~\cite{2023-ACL-RL4F} introduces a novel framework where the critic model is trained using RL, without modifying the parameters of the reasoning LLM. The critic model generates natural language feedback, which is then used by the LLM to correct its output. AutoMathCritique~\cite{2024-AutoMathCritique} follows a similar approach by decoupling the reasoning and criticism modules. It constructs various step-level error paths through methods like k-sample sampling, increasing temperature coefficients, and error insertion. These error paths are used to prompt GPT-4 to generate corresponding step-level critiques, which are then employed to automatically construct large-scale critique dataset. The trained critic model can then provide natural language feedback for mathematical reasoning tasks, thereby supervising the optimization of the actor model. As the reasoning iterations increase, the reasoning of the actor model is continuously improved, enhancing the performance in complex reasoning tasks.

In the field of multimodel, Critic-V~\cite{2024-critic-v} utilizes GPT-4 to artificially introduce errors into the collected VQA data, allowing multiple VLMs to generate critical comments for these errors. To evaluate the quality of the generated critiques, Rule-based Reward (RBR) combines the Jaccard index and GPT score to rank and score the critiques. Based on these preference-ranked data,  Direct Preference Optimization (DPO) is employed to train Critic-V, enabling it to generate higher-quality natural language feedback that offers more effective reasoning guidance for VQA tasks.
\section{Method}
\begin{figure*}[t]
	\centering
	\includegraphics[width=\textwidth]{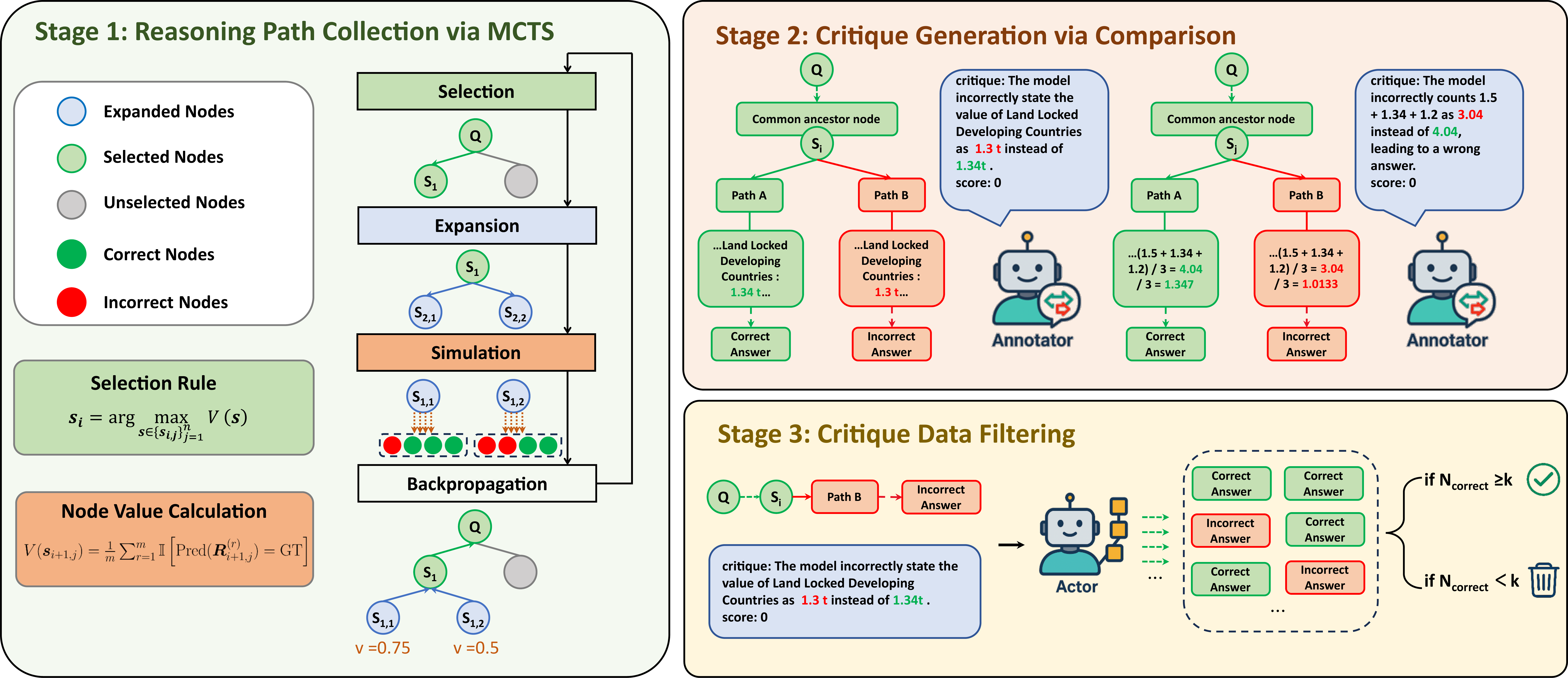}
	\caption{The construction pipeline of MMC dataset.}
	\label{fig:method}
\end{figure*}

As shown in Figure~\ref{fig:overview}, we build a multimodal actor-critic framework that consists of two key components: an actor model, which generates step-by-step reasoning paths conditioned on image-question inputs, and a critic model, which evaluates each reasoning steps and provides corrective feedback. The actor model iteratively refines its reasoning based on the feedback until the reasoning outcome is deemed satisfactory by the critic model. A key challenge in training the critic model lies in acquiring fine-grained supervision for critique quality without relying on costly manual annotations. To this end, we introduce an automated approach for constructing the MMC dataset as illustrated in Figure~\ref{fig:method}. More details are elaborated in the subsequent sections.
\subsection{Reasoning Path Collection with MCTS}
To capture the diverse types of reasoning errors that a VLM may encounter at each step, we use MCTS~\cite{2006-MCTS} to collect step-wise reasoning paths. This approach enables systematic exploration of possible reasoning paths, striking a balance between diversity (exploration) and quality (exploitation). We denote the actor model as $\pi_\theta$, initialized from a pretrained VLM. Given a multimodal question input $\bm{Q} = \{ \bm{x}, \bm{I} \}$—where $\bm{x}$ denotes the textual component (including a specially designed prompt to encourage step-by-step reasoning) and $\bm{I}$ represents the visual input—the actor model autoregressively generates a sequence of intermediate reasoning steps toward the final answer:

\begin{equation}(\bm{Q}, \bm{s}_1, \bm{s}_2, \bm{s}_3, \ldots, \bm{s}_M) \sim \pi_\theta(\cdot \mid \bm{Q}),\end{equation}
where each intermediate step $\bm{s}_i$ is constrained to a maximum of 30 tokens and serves as a fundamental unit for MCTS-based iterative exploration. In the tree structure constructed by MCTS, a node at the $i$-th level is represented as $\bm{s}_i = \{\mathcal{P}(\bm{s}_i), N(\bm{s}_i), V(\bm{s}_i)\}$, where $\mathcal{P}(\bm{s}_i)$ denotes the partial reasoning path from the root to node $\bm{s}_i$, $N(\bm{s}_i)$ is the visit count, and $V(\bm{s}_i)$ is the estimated value of the node. This structure facilitates fine-grained tracking of step-wise reasoning variations throughout the search process. To promote diversity while maintaining response quality, we apply a fixed temperature. Each iteration of MCTS consists of four key phases: \emph{selection}, \emph{expansion}, \emph{simulation}, and \emph{backpropagation}.

\noindent\textbf{Selection.}
The selection phase begins at the root node $\bm{Q}$, which we denote as the initial state $\bm{s}_0$, and recursively selects the next node $\bm{s}_i$ from the children of the current node $\bm{s}_{i-1}$ based on the highest estimated value:
\begin{equation}
	\bm{s}_i = \arg\max_{\bm{s} \in \{\bm{s}_{i,j}\}_{j=1}^n} V(\bm{s}),
\end{equation}
where $\{\bm{s}_{i,j}\}_{j=1}^n$ represents the set of $n$ candidate child nodes of $\bm{s}_{i-1}$, and $V(\bm{s})$ denotes the value estimate associated with node $\bm{s}$.

\noindent\textbf{Expansion.}
Once a leaf node $\bm{s}_i$ is selected during the selection phase, the expansion phase generates $n$ candidate next steps from the actor model $\pi_\theta$. Each candidate step is sampled from the current partial reasoning path $\mathcal{P}(\bm{s}_i)$, and truncated to a maximum length of $L_s$ tokens. This results in a set of child nodes:
\begin{equation}
	\{\bm{s}_{i+1,j}\}_{j=1}^{n} = \text{Truncate}_{\leq L_s}\left( \pi_\theta(\mathcal{P}(\bm{s}_i)) \right),
\end{equation}
where each $\bm{s}_{i+1,j}$ represents a step-level continuation. The sampled continuations are then added to the search tree as new child nodes of $\bm{s}_i$.

\noindent\textbf{Simulation.}
In the simulation phase, each newly expanded child node $\bm{s}_{i+1,j}$ is evaluated to obtain its estimated value $V(\bm{s}_{i+1,j})$. Specifically, starting from the partial reasoning path $\mathcal{P}(\bm{s}_{i+1,j})$, the autoregressively generates the remaining steps until an end-of-sequence token is reached or a predefined maximum token length is exceeded. This rollout process is repeated $m$ times under fixed sampling settings, resulting in a set of complete reasoning paths denoted as $\{\bm{R}^{(r)}_{i+1,j}\}_{r=1}^m$.
For each generated reasoning path, we compare its predicted answer with the ground-truth answer and assign a binary score of 1 if they match, and 0 otherwise. The estimated value of node $\bm{s}_{i+1,j}$ is then defined as the average score across the $m$ rollouts:
\begin{equation}
	V(\bm{s}_{i+1,j}) = \frac{1}{m} \sum_{r=1}^{m} \mathbb{I} \left[ \text{Pred}(\bm{R}^{(r)}_{i+1,j}) = \text{GT} \right],
\end{equation}
where $\mathbb{I}[\cdot]$ is the indicator function, $\text{Pred}(\cdot)$ denotes the predicted answer from the reasoning path, and $\text{GT}$ is the ground-truth answer.

\noindent\textbf{Backpropagation.}
After evaluating the expanded nodes, their estimated values are backpropagated along the traversal path to update the visit counts and value scores of the ancestor nodes $\bm{s}_j$ ($0 \leq j \leq i$). The updates are performed using the following equations:
\begin{align}
	N_{\text{new}}(\bm{s}_j) &= N_{\text{old}}(\bm{s}_j) + 1, \\
	V_{\text{new}}(\bm{s}_j) &= \frac{V_{\text{old}}(\bm{s}_j) \cdot N_{\text{old}}(\bm{s}_j) + V(\bm{s}_{i+1,j})}{N_{\text{new}}(\bm{s}_j)},
\end{align}
where $N_{\text{old}}(\bm{s}_j)$ and $V_{\text{old}}(\bm{s}_j)$ represent the previous visit count and value score of node $\bm{s}_j$ before backpropagation, and $V(\bm{s}_{i+1,j})$ is the value obtained from the simulation phase.

The iteration terminates when the selected node either includes an end-of-sequence token or exceeds a maximum token length.

\subsection{Critique Generation via Comparison}
To generate step-level critiques, we leverage the structural information in the reasoning tree constructed by MCTS, as illustrated in Stage~2 of Figure~\ref{fig:method}.

For each tree, we first select a high-quality reference reasoning path—specifically, a complete path that is ultimately selected under the MCTS policy and leads to a correct final answer. Then, for each reasoning path that results in an incorrect final answer, we identify its last common ancestor node with the reference path. This node corresponds to the latest reasoning state where both paths still agree. From that point onward, the incorrect path deviates from the correct one; we denote the correct branch as path A and the incorrect branch as path B.

Benefiting from the step-wise reasoning paths constructed via MCTS, our framework only requires the annotator model to compare path A and path B to generate a natural language critique. This localized comparison substantially lowers the difficulty of critique generation, allowing even relatively weak models to produce meaningful feedback. To prevent the critique data from distilling reasoning outcomes from more powerful models such as GPT-4o~\cite{2024-critic-v}, we adopt a self-annotation setup in which the actor model itself serves as the annotator. Rather than directly answering the original question, the annotator is instructed to focus solely on identifying and explaining the specific reasoning mistake. The goal is to produce targeted corrective feedback that guides the actor model to iteratively refine its reasoning path toward the correct one.


 Based on this process, we construct MMC dataset where each sample consists of a multimodal question input, a reasoning path, a binary correctness score, and a critique text, denoted as $(\bm{Q}, \bm{A}, v, \bm{C})$. For reasoning paths that lead to the correct final answer, we standardize the critique text as ``No corrections needed.'' to serve as positive examples.
 More details can be found in the Appendix.

\subsection{Critique Data Filtering}

To ensure the quality and usefulness of generated critiques, we adopt an automatic filtering strategy based on their ability to help the actor model correct its own reasoning.

As shown in Stage 3 of Figure~\ref{fig:method}, for each negative sample, we prompt the actor model to refine the original incorrect reasoning path using the corresponding critique text. Specifically, the actor model takes as input the triplet $(\bm{Q}, \bm{A}, \bm{C})$ and generates $10$ refined answers under fixed sampling settings. Each refined answer is then compared against the ground-truth answer to determine whether the prediction is correct. If the number of correct refinements is greater than or equal to a predefined threshold $K=3$, the critique sample is considered effective and retained in the dataset; otherwise, it is discarded.

\subsection{Training the Critic Model}
 We fine-tune the critic model $\pi_\phi$ using the MMC dataset.
 
 \noindent\textbf{Language Modeling Loss.}  
 To supervise critique generation, we employ a language modeling head trained with the standard cross-entropy loss~\cite{2023-ICML-CELoss}: 
 \begin{equation}
 	\mathcal{L}_{\text{lm}} = -\sum_{t=1}^{T} \log P_\phi(c_t \mid Q, A, c_{<t}),
 \end{equation}
 where $c_t$ is the $t$-th token in the target critique $C$, and $\phi$ denotes the parameters of the critic model.
 
\noindent\textbf{Score Prediction Loss.}  
To enable our critic model to better assess the quality of responses generated by the actor model, we incorporate a design inspired by the Outcome Reward Model (ORM)\cite{ORM}. Specifically, we append a score head to the critic model, implemented as a multi-layer perceptron (MLP) that outputs a scalar for each token. The scalar prediction at the last token is used as the estimated correctness of a given response. We supervise the score prediction using a binary cross-entropy loss\cite{2023-ICML-CELoss}:


\begin{equation}
	\mathcal{L}_{\text{score}} = - \left[ v \log \hat{v} + (1 - v) \log (1 - \hat{v}) \right],
\end{equation}
where $\hat{v} \in [0, 1]$ denotes the predicted scalar score and $v \in \{0, 1\}$ is the ground-truth label.
%
%
%

\noindent\textbf{Overall Objective.}
The total training objective is a weighted sum of the language modeling loss and the score prediction loss: \begin{equation} \mathcal{L}_{\text{total}} = \mathcal{L}_{\text{lm}} + \lambda \mathcal{L}_{\text{score}}, \end{equation} where $\lambda$ is a hyperparameter balancing the two loss terms.

\subsection{Iterative Inference with the Actor–Critic Framework}
%

\begin{figure}[t] 
	\centering \includegraphics[width=\columnwidth]{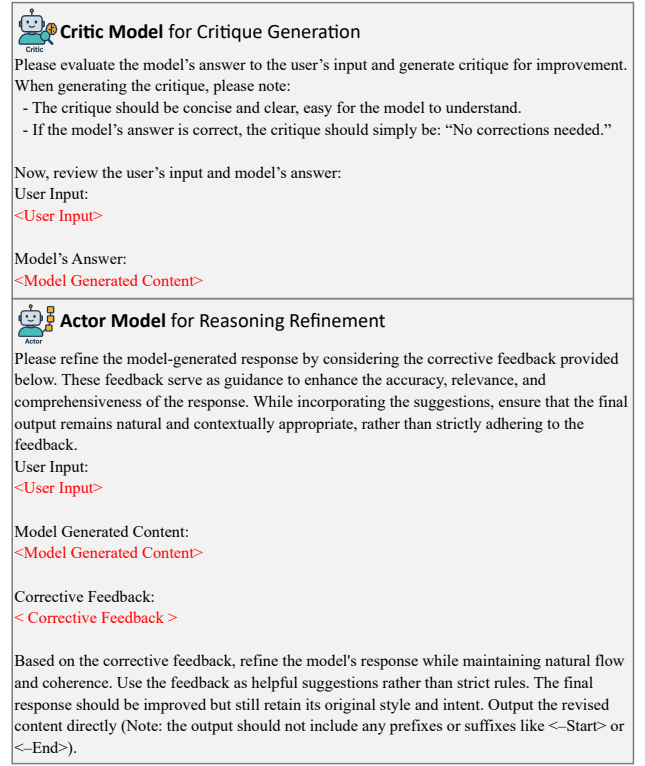} 
	\caption{Prompt templates used for actor and critic during inference.} \label{fig:actor-critic-prompt} 
\end{figure}

At inference time, we adopt an iterative actor–critic framework, where the actor model progressively refines its output based on feedback from the critic. The process begins with the actor generating a step-by-step reasoning. At each iteration, the critic evaluates the reasoning by assigning a scalar score $v$ and producing a natural language critique $\bm{C}$. The actor then updates its reasoning accordingly. The prompt templates used for the actor and the critic are illustrated in Figure~\ref{fig:actor-critic-prompt}.

This iterative process continues until the answer is deemed satisfactory by the critic model, i.e., when the scalar score $\sigma$ exceeds a predefined threshold $\gamma$.
We formalize this iterative procedure in Algorithm~\ref{alg:actor_critic}.

\begin{algorithm}[h]
	\caption{Iterative Actor–Critic Inference}
	\label{alg:actor_critic}
	\KwIn{Multimodal question $\bm{Q} = \{\bm{x}, \bm{I}\}$, actor model $\pi_\theta$, critic model $\pi_\phi$, threshold $\gamma$, maximum steps $T$}
	\KwOut{Final refined answer $\bm{A}_t$}
	
	Initialize answer $\bm{A}_1 \leftarrow \pi_\theta(\bm{Q})$\;
	
	\For{$t = 1$ \KwTo $T$}{
		Evaluate with critic: $(\bm{C}_t, v_t) \leftarrow \pi_\phi(\bm{Q}, \bm{A}_t)$\;
		
		\If{$v_t > \gamma$}{
			\textbf{break}\;
		}
		
		Refine with feedback: $\bm{A}_{t+1} \leftarrow \pi_\theta(\bm{Q}, \bm{A}_t, \bm{C}_t)$\;
	}
	\Return $\bm{A}_t$\;
\end{algorithm}

\section{Experiments}
\begin{table*}[ht]
	
	\centering
		\caption{Comparison across benchmarks. We group models into three categories: large-scale or proprietary models, open-source models of comparable scale, and our open-source models with critic. }
	\small
	\begin{tabular}{lccccccc}
		\toprule
		\textbf{Model} & \textbf{MMStar}~\cite{MMStar} & \textbf{ScienceQA}~\cite{ScienceQA} & \textbf{ChartQA}~\cite{2022-ACL-ChartQA} & \textbf{MathVista}~\cite{MathVista} & \textbf{MathVision}~\cite{MathVision} & \textbf{MathVerse}~\cite{MathVerse} & \textbf{M\({}^{\mbox{3}}\)CoT}~\cite{2024-m3cot} \\
		\midrule
		\multicolumn{8}{c}{\textit{Proprietary and Large-Scale Open-Source Models}} \\
		\midrule
		GPT-4o~\cite{2024-gpt4o} & 63.9 & - & 85.7 & 63.8 & 30.4 & 50.2 & 64.3 \\
		Claude-3.5 Sonnet~\cite{claude3.5sonnet2024} & 65.1 & - & 90.8 & 67.7 & 35.6 & - & - \\
		Qwen2.5-VL-72B~\cite{Qwen2.5-VL} & 70.8 & - & 89.5 & 74.8 & 38.1 & 57.6 & - \\
		Qwen2-VL-72B~\cite{Qwen2-VL} & 68.3 & - & 88.3 & 70.5 & 25.9 & - & - \\
		InternVL2-Llama3-76B~\cite{InternVL} & 67.4 & - & 88.4 & 65.6 & 33.6 & - & - \\
		\midrule
		\multicolumn{8}{c}{\textit{Open-Source Models of Comparable Scale}} \\
		\midrule
		DeepSeek-VL2-MOE-4.5B~\cite{DeepSeek-VL2} & 61.3 & - & 86.0 & 62.8 & - & - & - \\
		MiniCPM-V-2.6-8B~\cite{MiniCPM-V} & 57.5 & 90.9 & - & 60.6 & - & 24.1 & 56.0 \\
		LLaVA-OneVision-7B~\cite{LLaVA-OneVision} & 61.7 & \textbf{96.0} & 80.0 & 63.2 & 18.4 & 26.2 & 52.3 \\
		\midrule
		\multicolumn{8}{c}{\textit{Open-Source Models with Critic (Ours)}} \\
		\midrule
		Qwen2-VL-7B~\cite{Qwen2-VL} & 60.7 & 80.1 & 83.0 & 58.2 & 16.3 & 31.9 & 57.8 \\
		\textbf{Qwen2-VL-7B + Critic} & 60.9\supgain{0.2} & 91.7\supgain{11.6} & 85.9\supgain{2.9} & 68.1\supgain{9.9} & 22.4\supgain{6.1} & 39.0\supgain{7.1} & 73.7\supgain{15.9} \\
		\midrule
		Qwen2.5-VL-7B~\cite{Qwen2.5-VL} & 63.9 & 81.6 & 87.3 & 68.2 & 25.1 & 49.2 & 67.6 \\
		\textbf{Qwen2.5-VL-7B + Critic} & \textbf{64.4}\supgain{0.5} & 92.1\supgain{10.5} & \textbf{89.1}\supgain{1.8} & \textbf{73.6}\supgain{5.4} & \textbf{27.5}\supgain{2.4} & \textbf{49.6}\supgain{0.4} & \textbf{76.2}\supgain{8.6} \\
		\midrule
		InternVL2-8B~\cite{InternVL} & 61.5 & 88.4 & 83.3 & 58.3 & 18.4 & 37.0 & 57.6 \\
		\textbf{InternVL2-8B + Critic} & 62.1\supgain{0.6} & 93.3\supgain{4.9} & 85.5\supgain{2.2} & 67.3\supgain{9.0} & 23.7\supgain{5.3} & 41.9\supgain{4.9} & 70.6\supgain{13.0} \\
		
		\bottomrule
	\end{tabular}

	\label{tab:full-benchmark}
\end{table*}

\subsection{Experimental Setup}

\noindent\textbf{Benchmarks and Metrics.}
We evaluate our method on a diverse set of multimodal reasoning benchmarks, covering both general VQA and mathematical reasoning tasks. Specifically, we include (1) general VQA: ChartQA~\cite{2022-ACL-ChartQA}, ScienceQA~\cite{ScienceQA}, and MMStar~\cite{MMStar}; and (2) mathematical reasoning: MathVista~\cite{MathVista}, MathVerse~\cite{MathVerse}, MathVision~\cite{MathVision}, and M\({}^{\mbox{3}}\)CoT~\cite{2024-m3cot}.

To assess answer correctness, we use GPT-4o~\cite{2024-gpt4o} to compare the predictions generated by the model against a ground truth.

\noindent\textbf{Implementation details.}
During the construction of the MMC dataset, we employ Qwen2-VL-7B~\cite{Qwen2-VL} and Qwen2.5-VL-7B~\cite{Qwen2.5-VL} as actor models, using a decoding temperature of 0.7. When serving as the annotator model, the decoding temperature is set to 0. The sampled queries are drawn from multiple VQA datasets, including MathV360k~\cite{2024-EMNLP-Math-LlaVA} as well as the training sets of  ChartQA~\cite{2022-ACL-ChartQA} and M\({}^{\mbox{3}}\)CoT~\cite{2024-m3cot}, which share the same data sources as the training sets of our downstream evaluation benchmarks. To improve annotation reliability, we exclude most multiple-choice and true/false questions, as they often introduce inconsistencies.
We fine-tune Qwen2-VL-7B as the critic model using the MMC dataset, with a fixed learning rate of $1 \times 10^{-5}$ and the AdamW optimizer. During inference, we use Qwen2-VL-7B, Qwen2.5-VL-7B, and InternVL2-8B~\cite{InternVL} as actor models, all with a decoding temperature of 0.7. The critic model adopts greedy decoding.
The maximum number of actor-critic iterations is set to $T=5$. All experiments are conducted on Tesla A800 GPUs.

\subsection{Results}

Table~\ref{tab:full-benchmark} presents the performance comparison of our method with a variety of VLMs across various reasoning benchmarks. 

\noindent\textbf{Comparison with baselines.}
We compare the performance of our chosen actor models with and without the critic across seven multimodal reasoning benchmarks. As shown in Table~\ref{tab:full-benchmark}, incorporating the critic consistently improves performance across all tasks. For instance, Qwen2-VL-7B with the critic achieves gains of +11.6 on ScienceQA, +9.9 on MathVista, +7.1 on MathVerse, and +15.9 on M\({}^{\mbox{3}}\)CoT. Similarly, Qwen2.5-VL-7B with the critic yields gains of +10.5, +5.4, +0.4, and +8.6, respectively. InternVL2-8B also benefits substantially, with improvements such as +9.0 on MathVista and +13.0 on M\({}^{\mbox{3}}\)CoT. It is worth noting that the MMC dataset used to train the critic model does not contain any samples derived from InternVL2-8B, and the critic is consistently trained with Qwen2-VL-7B as the base model. Despite this, InternVL2-8B still exhibits clear gains when paired with the critic, highlighting the generalization ability of our feedback mechanism across different actor backbones. Furthermore, all of our critic-augmented models outperform strong open-source models of comparable scale in complex mathematical reasoning tasks , demonstrating the effectiveness and robustness of our approach in enhancing reasoning quality.

\noindent\textbf{Comparison with SOTA.}
We also compare our actor-critic framework against a range of proprietary and large-scale open-source models. Despite being built upon relatively small base models, our approach achieves competitive performance and even surpasses some larger-scale counterparts on several benchmarks by increasing inference-time computation through iterative feedback. These results demonstrate the effectiveness of our actor-critic framework in enhancing the reasoning capabilities of VLMs, offering an  alternative for complex multimodal reasoning tasks.

\begin{table*}[t]
	\centering
	\caption{Comparison of different critics on four multimodal reasoning benchmarks.}
	\begin{tabular}{lcccc}
		\toprule
		\textbf{Model} & \textbf{MMStar} & \textbf{ScienceQA} & \textbf{MathVista} & \textbf{MathVerse} \\
		\midrule
		Qwen2-VL-7B          & 60.7 & 80.1 & 58.2 & 31.9 \\
		\quad + Critic-V~\cite{2024-critic-v}     & 56.2\suploss{4.5} & 74.5\suploss{5.6} & \textbf{73.2}\supgain{15.0} & 32.9\supgain{1.0} \\
		\quad + \textbf{Our Critic} & \textbf{60.9}\supgain{0.2} & \textbf{91.7}\supgain{11.6} & 68.1\supgain{9.9} & \textbf{39.0}\supgain{7.1} \\
		\bottomrule
	\end{tabular}
	
	\label{tab:our-critic}
\end{table*}

\begin{figure*}[t]
	\centering
	\includegraphics[width=\textwidth]{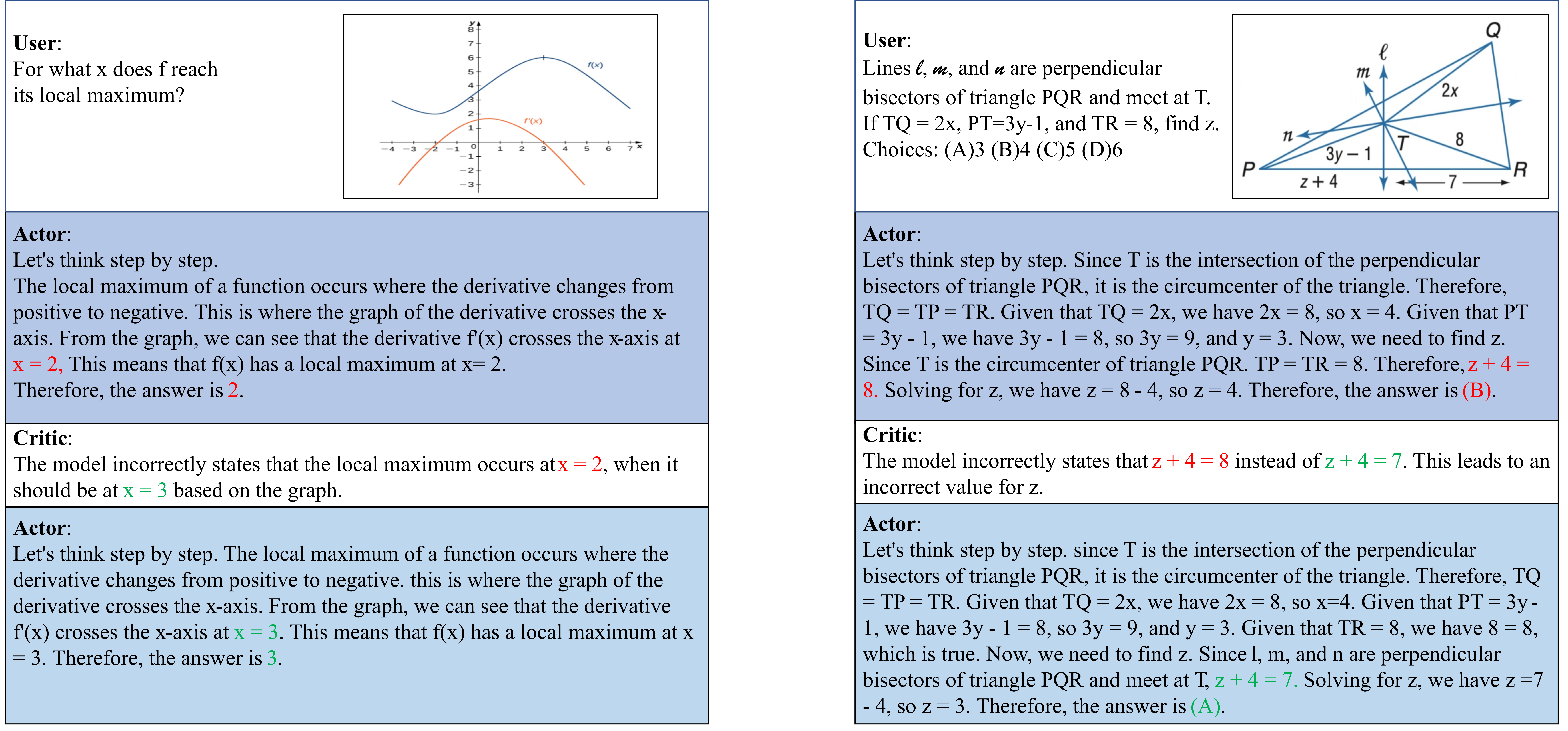}
	\caption{Case studies on evaluation samples from MathVista. Our critic model is capable of identifying both visual perception error (left) and reasoning error (right), and provides corrective feedback to guide the actor model in refining its reasoning and arriving at the correct answer.}
	\label{fig:good-case}
\end{figure*}
\sloppy
\noindent\textbf{Comparison with Recent Method.}
We compare our multimodal critic model with the recent Critic-V~\cite{2024-critic-v} across four benchmarks. As shown in Table~\ref{tab:our-critic}, our model outperforms Critic-V on three benchmarks, notably ScienceQA and MathVerse, while underperforming on MathVista. In Critic-V’s setup, the actor model is prompted to provide short answers—typically single words or phrases and the critic model often reveals the correct answer directly within its feedback. As a result, the refinement process by the actor model tends to resemble copying the final answer rather than true reasoning refinement.
In contrast, our actor model is required to generate step-by-step reasoning, and our critic model is specifically designed to identify and comment on reasoning errors without explicitly revealing the answer. As illustrated in Figure~\ref{fig:good-case}, our critic model is capable of detecting both visual perception error (left) and reasoning mistake (right) without providing the final answer directly. This enables the actor model to refine its reasoning based on the feedback and ultimately arrive at the correct answer.

\subsection{Ablation Study}  
To better understand the contribution of each component in our framework, we conduct an ablation study using Qwen2-VL-7B as the actor model on the MathVista dataset. For the critic model, we evaluate four variants T1 through T4, starting from a baseline without any specialized training. Key components—including fine-tuning on the MMC dataset, critique data filtering, and the use of score head—are added incrementally across the variants. This step-by-step comparison allows us to isolate and quantify the effect of each module on the critic’s ability to generate accurate and helpful corrective feedback.
\begin{table}[t]
	\centering
		\caption{Results of the evaluation on the MathVista dataset. ``MMC” denotes the use of the MMC dataset. ``DF'' indicates the inclusion of the data filtering, and ``SH” signifies the use of score head.  ``Acc.” corresponds to the performance.}
	\renewcommand{\arraystretch}{0.7}  
	\begin{tabular}{cccccc}
		\toprule  
		Variants & MMC & DF & SH &Acc. \\  
		\midrule  
		T1 & \ding{56} & \ding{56} & \ding{56}& 58.9 \\
		T2 & \ding{52} & \ding{56} & \ding{56}& 63.3 \\
		T3 & \ding{52} & \ding{52} & \ding{56}& 64.8 \\
		T4 & \ding{52} & \ding{52} & \ding{52}& 68.1 \\
		\bottomrule  
	\end{tabular}

	\label{table:ablation}
\end{table}

\noindent\textbf{Effectiveness of MMC.}  
The MMC dataset is specifically constructed to enhance the critic model’s ability to detect reasoning errors and provide targeted corrective feedback. We evaluate its effectiveness by comparing two settings: T1, where the critic model is directly initialized from the actor model without further training, and T2, where the critic is fine-tuned on the MMC dataset. As shown in Table~\ref{table:ablation}, T2 significantly outperforms T1, demonstrating that training on MMC substantially improves the critic’s capacity to identify reasoning flaws and generate precise feedback.

\noindent\textbf{Effectiveness of Data Filtering.}  
We hypothesize that high-quality critiques are those that effectively guide the actor model from an incorrect reasoning path to the correct one, ultimately leading to the right answer. Based on this intuition, we apply a filtering strategy to remove low-quality critiques from the training data. Compared to T2, T3 trains the critic model on a filtered subset of the data. The resulting performance gain demonstrates that data filtering helps the critic model learn to generate more effective critiques.

\noindent\textbf{Effectiveness of Score Head.}
To enhance the critic model’s ability to evaluate answer quality, we introduce a score head inspired by the Outcome Reward Model (ORM)~\cite{ORM}, which predicts the correctness of the actor model’s response. This design provides auxiliary supervision during training, encouraging the critic to align critique generation with answer verification. At inference time, the predicted score serves as an internal signal to assess whether the actor’s response requires refinement. The T4 variant, which incorporates this component on top of T3, achieves the best performance among all configurations, improving accuracy from 64.8\% to 68.1\%. These results confirm the effectiveness of the score head in improving both critique quality and evaluation accuracy.

\begin{figure}[t]
	\centering
	\includegraphics[width=\columnwidth]{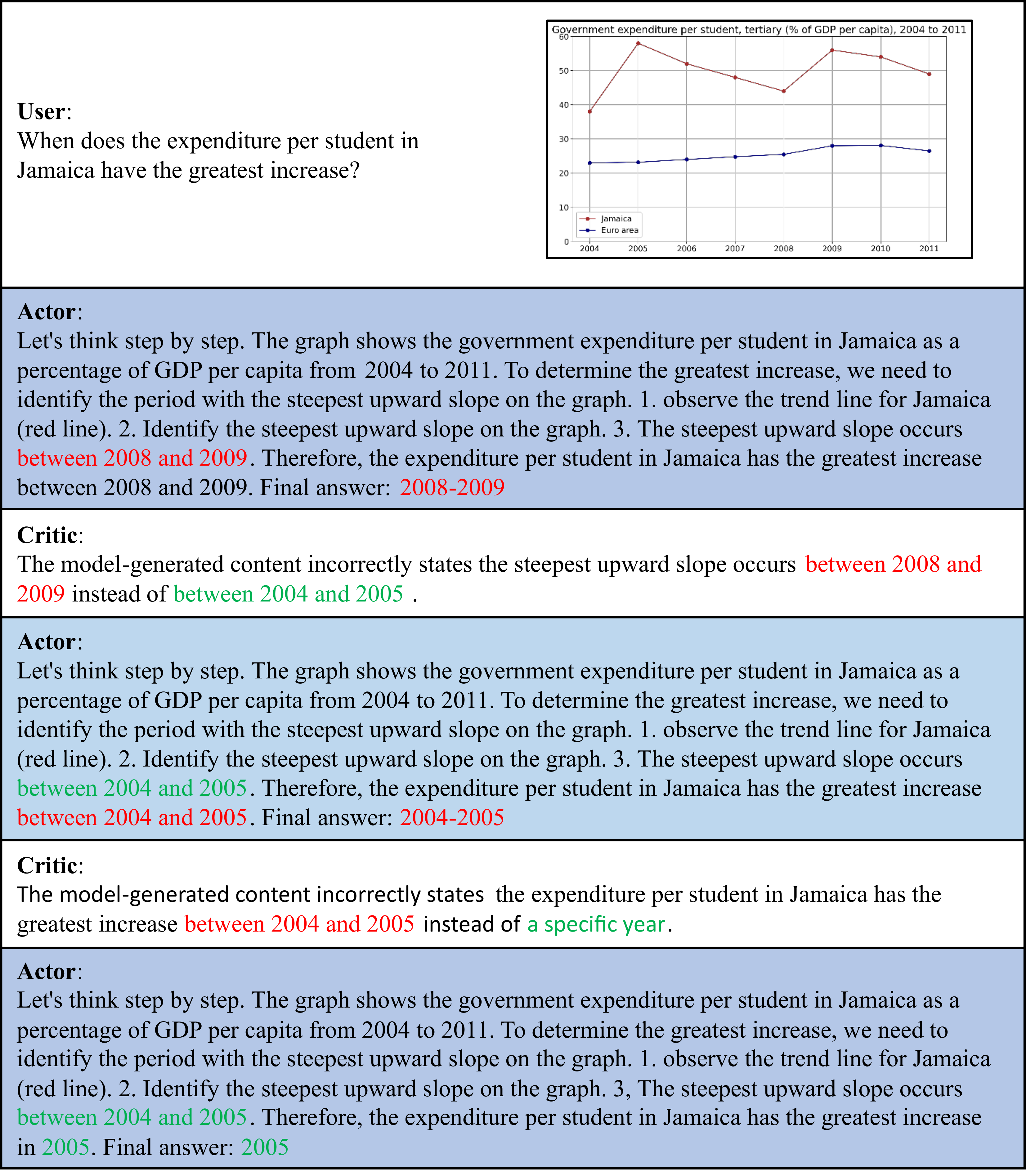}
	\caption{An example of iterative refinement. The actor model produces an initially incorrect answer and iteratively corrects its reasoning through critic's feedback, ultimately converging to the correct solution after two iterations.}
	\label{fig:multiiter}
\end{figure}
\noindent\textbf{Effectiveness of Iteration.}
To evaluate the impact of iterative refinement, we analyze the actor–critic interaction process on the MathVista dataset over multiple iterations. As shown in Table~\ref{tab:iteration}, the answer accuracy improves steadily from 58.2\% at iteration 0 (without feedback) to 68.1\% after four refinement steps.
At the same time, the number of responses identified by the critic model as requiring further refinement—denoted as $N_{\text{refine}}$—decreases from 335 in the first iteration to 115 in the final one. This trend demonstrates that our critic provides effective corrective feedback, guiding the actor to iterately correct its reasoning toward correct answer. Figure~\ref{fig:multiiter} illustrates a case study where the actor model arrives at the correct answer after two rounds of refinement.

\begin{table}[h]
	\centering
	\caption{Effect of iterative refinement on MathVista. Accuracy improves with each actor–critic iteration, while the number of refined reasoning decreases.}
	\label{tab:iteration}
	\begin{tabular}{c|ccccc}
		\toprule
		\textbf{iter} & \textbf{0} & \textbf{1} & \textbf{2} & \textbf{3} & \textbf{4} \\
		\midrule
		Acc. & 58.2 & 66.0 & 67.5 & 67.9 & 68.1 \\
		$N_{\text{refine}}$ & - & 335 & 170 & 137 & 115 \\
		\bottomrule
	\end{tabular}
\end{table}

\noindent\textbf{Impact of Instruction Following of Actor.}
We further conduct experiments using Qwen2-VL-2B as the actor model. As shown in Table~\ref{tab:critic-improvement}, although Qwen2-VL-2B achieves a lower baseline performance, its improvement after incorporating the critic is noticeably smaller than that of Qwen2-VL-7B. This suggests that even when the critic model provides accurate critiques, the effectiveness of refining the reasoning toward the correct answer is still constrained by the actor model’s ability to follow instructions. We provide a detailed error analysis in the Appendix.

\begin{table}[ht]
	\centering
	\caption{Performance comparison between Qwen2-VL-7B and Qwen2-VL-2B with and without critic model.}
	\begin{tabular}{l|cccc}
		\toprule
		\textbf{Model}  & \textbf{MathVision} & \textbf{MathVerse} & \textbf{M\({}^{\mbox{3}}\)CoT} \\
		\midrule
		Qwen2-VL-7B  & 16.3 & 31.9 & 57.8 \\
		\textbf{ + Critic}   &22.4\supgain{6.1} & 39.0\supgain{7.1} & 73.7\supgain{15.9} \\
		\midrule
		Qwen2-VL-2B   & 12.4 & 21.0 & 46.7 \\
		\textbf{ + Critic}   &14.6\supgain{2.2} &21.5\supgain{0.5} & 55.4\supgain{8.7} \\
		\bottomrule
	\end{tabular}
	
	\label{tab:critic-improvement}
\end{table}

\section{Conclusion}

In this work, we present a multimodal actor–critic framework that enhances the reasoning of VLMs through iterative feedback and refinement. The actor model generates step-by-step reasoning paths conditioned on visual and textual inputs, while the critic model evaluates these paths and provides targeted corrective feedback. To alleviate the need for costly manual supervision, we introduce an automated method for constructing the multimodal critique dataset. By leveraging MCTS, we systematically explore diverse reasoning paths, and generate critiques by prompting an annotator model to compare locally divergence. This enables the construction of MMC dataset, built upon which our training and inference pipeline supports a generalizable and scalable actor–critic paradigm. Extensive experiments across multiple public benchmarks and mainstream VLMs demonstrate that our approach significantly improves VLM performance on complex multimodal reasoning tasks. These results highlight the effectiveness and broad applicability of feedback-driven reasoning refinement for VLMs.

\bibliographystyle{ACM-Reference-Format}
\bibliography{sample-base}


%
%
%
%
%
%
%
\clearpage
\appendix
\section{MMC Dataset}
To generate step-level critiques, we prompt an annotator model to compare pairs of reasoning paths diverging from a shared ancestor node—one leading to a correct conclusion and the other to an incorrect one.
The prompt template is shown in Figure~\ref{fig:annotator}.
Based on this process, we construct MMC dataset where each sample consists of a multimodal question input, a reasoning path, a binary correctness score, and a critique text, denoted as $(\bm{Q}, \bm{A}, v, \bm{C})$. For reasoning paths that lead to the correct final answer, we standardize the critique text as ``No corrections needed.'' to serve as positive examples. 
We plan to release the full MMC dataset in future work. 

\begin{figure*}[h] 
	\centering 
	\includegraphics[ width=0.9\textwidth]{Annotator.pdf} 
	\caption{Prompt template used for annotator to generate critique.} \label{fig:annotator} 
\end{figure*}
\begin{figure*}[t] 
	\centering 
	\includegraphics[ width=0.9\textwidth]{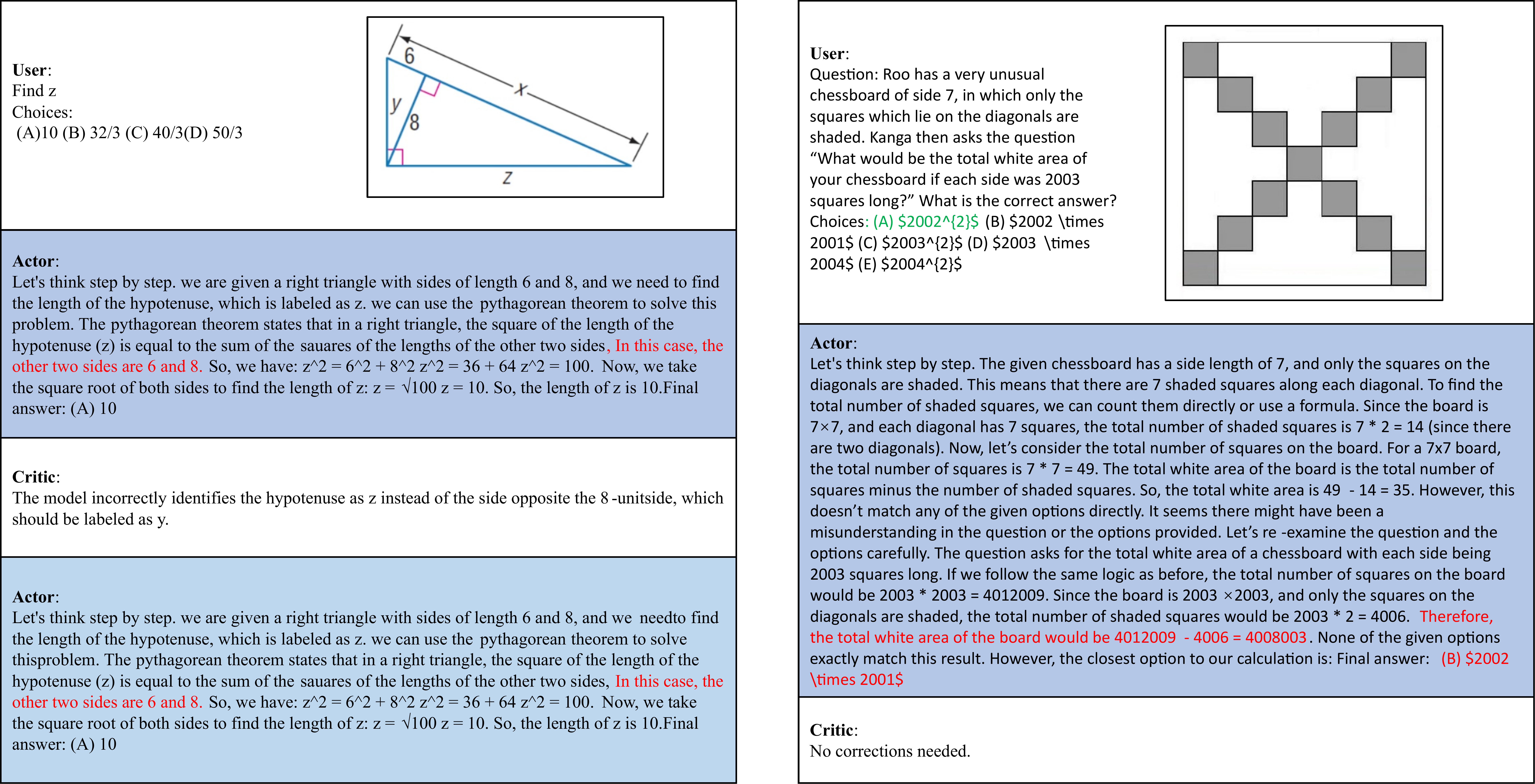} 
	\caption{Examples of failure cases: the actor fails to follow feedback (left), and the critic fails to detect subtle errors in a complex reasoning task (right).} \label{fig:badcase} 
\end{figure*}

\section{Error Analysis}
Despite being fine-tuned to identify reasoning errors and provide corrective feedback, our critic model still faces two major challenges.

First, even when the critic accurately identifies flaws in the actor model’s reasoning, the success of refinement ultimately depends on the actor model’s ability to follow instructions. As illustrated in the left example of Figure~\ref{fig:badcase}, the critic correctly points out that the actor incorrectly identifies the hypotenuse as z instead of the side opposite the 8-unitside, which should be labeled as y. However, the actor fails to refine its reasoning accordingly, resulting in an incorrect final answer.

Second, while the critic model improves the actor's stability in basic capabilities such as recognition and computation, it remains limited in handling more complex analytical tasks. These cases demand stronger intrinsic reasoning ability to identify and validate subtle logical flaws. As shown in the right example of Figure~\ref{fig:badcase}, the critic fails to detect a nuanced mistake—specifically, the double-counting of the overlapping cell at the intersection of diagonals.

Addressing these two limitations will be the focus of our future work.
\end{document}